\documentclass[a4paper]{jpconf}
\usepackage{graphicx}

\def\solar{\ifmmode_{\mathord\odot}\else$_{\mathord\odot}$\fi~}
\def\gsim{\stackrel{>}{_\sim}}
\def\lsim{\stackrel{<}{_\sim}}

\begin{document}

\title{Sub-milliarcsecond Imaging of Sgr\,A* and M\,87}

\author{
T.P. Krichbaum$^1$, D.A. Graham$^1$, M. Bremer$^2$, W. Alef$^1$, A. Witzel$^1$, J.A. Zensus$^1$, and A. Eckart$^3$
}

\address{$^1$ MPIfR, Auf dem H\"ugel 69, 53121 Bonn, Germany}
\address{$^2$ IRAM, 300 rue de la Piscine, 38406 St. Martin d'Heres, France}
\address{$^3$ University of Cologne, I. Physikalisches Institut, Z\"ulpicher Str. 77, 50937 K\"oln, Germany}

\ead{tkrichbaum@mpifr-bonn.mpg.de}

\begin{abstract}
We present and discuss new result from mm-VLBI observations of M\,87 and Sgr\,A*. The imaging of these sources with a spatial
resolutions of a few to a few ten Schwarzschild radii offers new possibilities to study the immediate environment of
super-massive black holes.
\end{abstract}

\vspace{-0.8cm}
\section{Introduction}
In astronomy, the highest angular resolution is obtained with \underline{V}ery \underline{L}ong \underline{B}aseline 
\underline{I}nterferometry (VLBI). VLBI observations at short millimeter wavelengths (mm-VLBI) provide angular resolutions
in the ten micro-arcsecond range (i.e. $14$\,$\mu$as for a $10,000$\,km baseline at $\lambda=1.3$\,mm ($\nu=230$\,GHz)).
While VLBI observations at 2\,mm and 1\,mm wavelength are still in a stage of technical development, 3\,mm VLBI
experiments are performed on a regular basis.

The Global mm-VLBI Array (GMVA\footnote{web link: http://www.mpifr-bonn.mpg.de/old\_mpifr/div/vlbi/globalmm})
is operational since early 2000.
At 86\,GHz ($\lambda=3.5$\,mm), it combines the European antennas (Effelsberg 100\,m, Pico Veleta 30\,m, 
the phased Plateau de Bure interferometer 6x15\,m,
Onsala 20\,m, Mets\"ahovi 14\,m) with the VLBA. With the participation of the two sensitive IRAM telescopes and the 100\,m
Effelsberg telescope, the array sensitivity is improved by a factor of $3-4$, when compared to the VLBA alone.
For compact galactic and extragalactic radio sources, the GMVA provides VLBI images with an angular resolution of down to
$40$\,$\mu$as.

In the following we present and discuss results from mm-VLBI observations of the nuclei of M\,87 and of Sgr\,A*.
When expressed in terms of Schwarzschild radii ($R_{S} = 2 G M_{\rm BH} / c^2$), the spatial resolution
obtained for M\,87 and Sgr\,A* become very comparable. M\,87 is about $\sim 10^3$ times more distant
than Sgr\,A*, but has a putative black hole at its center, which is also $\sim 10^3$ times more massive than
that of Sgr\,A*. Present day 3\,mm-VLBI therefore allows to image both sources with a spatial resolution of 
only $10-20 R_{S}$. In view of the morphological difference between Sgr\,A* and M\,87 
(the latter exhibits a long relativistic jet), a comparison of the sub-milliarcsecond structure and kinematics of both
objects may lead to a better understanding of the astro-physical processes acting in the vicinity of such
super-massive black holes.

\section{3\,mm VLBI observations of M\,87}

The giant elliptical galaxy M\,87 (NGC\,4486, 3C\,274, Virgo\,A) is one of the closest radio galaxies
with a prominent radio jet. Its relative brightness at millimeter wavelengths makes it ideally suited 
for high angular resolution mm-VLBI studies, which address the still unanswered question how the powerful 
jets in radio-galaxies and quasars are launched and collimated.

At a distance of D=16.75\,Mpc (Whitmore et al. 199) an angular scale of 0.1\,mas
corresponds to a spatial scale of $2.5 \cdot 10^{16}$\,cm (0.0081\,pc). This translates to $27.8 R_{S}$ Schwarzschild radii 
for an assumed $3 \cdot 10^9 M\solar$ super-massive black hole at the center of M\,87 (e.g. Machetto et al. 1997).
Global 3\,mm VLBI observations, performed at the edge of the accessible VLBI observing bands, have
revealed new images of the inner-most region of M\,87. The 3\,mm VLBI images shown in Figure 1 were made from
observations performed during 2002--2004 and show a one-sided core-jet structure, extending $\sim 2$\,mas to the
west. The complexity of the source structure and the only moderate dynamic range of the maps (lower sensitivity in earlier
observations) do not yet allow to relate the individual structural components seen at the different epochs
with each other and by this determine the jet kinematics. Despite the obvious limitations, which will be overcome
in the near future, a number of basic statements can already be made:

\noindent
(i) In the maps, an unresolved compact component at the eastern end of the jet is seen. An upper limit to its
size is given by the (elliptical) VLBI observing beam. The uniformly weighted and un-tapered beam size of the
observations of April 2004 is $(193 \times 54) \mu$as, which gives an upper limit to the
size of the jet base (VLBI core) of $(54 \times 15) R_{S}$. Circular Gaussian model fitting 
yields a typical flux density of $\sim 0.5$\,Jy and a size of $\lsim 50 \mu$as ($\lsim 14 R_{S}$) for the VLBI core.
This results in a brightness temperature of $\gsim 10^{10.5}$\,K, clearly indicative of non-thermal (synchrotron) 
radiation. 

\noindent
(ii) The 86\,GHz VLBI map of April 2004 looks remarkably similar to two previously observed 43\,GHz VLBI maps,
which indicate conical jet-expansion on sub-milliarcsecond scales (Junor et al. 1999, Ly et al. 2004). The new
86\,GHz maps, which have a $\sim 2$ times higher resolution, reveal a bifurcated (possibly edge-brightened) 
expanding one-sided jet, with the southern edge of the emission cone being brighter and more compact 
than the northern part. The fact that the northern edge of the cone (at $r \leq 0.5$\,mas) is almost invisible in the maps 
of 2002 and 2003 may be explained with dynamic range limitations in these two maps.

\noindent
(iii) The large morphological differences seen between the three 86\,GHz maps indicate structural variations
on time-scales of less than one year.
On angular scales $\leq 0.5$\,mas, all three maps show some south-west oriented jet emission oriented along a position angle 
of $\sim -110^\circ$. In April 2003, the source was $\sim 1.5-2$ times brighter than usual, and a bright component
was seen at the map center (at zero position). Since in hybrid VLBI imaging using closure phases the absolute 
position information is lost, this bright component most likely should not be identified with the jet base (VLBI-core).
Instead, the fainter component located $\sim 0.2$\,mas in the north-east, may be identified with the VLBI-core. 
We note the tentative detection of a possible counter-jet at 43\,GHz (Ly et al. 2004), which however was
not seen earlier (Junor et al. 1999), nor at longer wavelengths (Dodson et al. 2006), 
nor in our 86\,GHz VLBI maps of 2002 and 2004. 

\noindent
(iv) Owing to the limited time sampling at 86\,GHz, a detailed kinematic study of the component motion in the inner
jet of M\,87 is not yet possible. Similar to the case of many other VLBI jets, we may assume that the bright central component 
seen in April 2003 was ejected from the VLBI core after a flux density outburst. This would give an apparent rate of 
change of $\gsim 0.2$\,mas/yr, corresponding to an apparent velocity of $\gsim 0.05$\,c. This speed would be 
consistent with the observed motion of $0.04 \pm 0.02$\,c seen at larger core separations with 
VLBI at 15\,GHz (Kellermann et al. 2004), and with a lower limit of 0.25\,c from VLBI at 43\,GHz (Ly et al. 2004b).
On the other hand, if we compare the relative positions of the two main secondary components seen in April 2002 
and April 2004 (at $r \simeq 0.25$\,mas \& $0.50$\,mas, respectively) stationarity of the inner jet components cannot be excluded. 
More frequent VLBI observations are necessary to determine the jet kinematics.

We like to point out that a measurement of VLBI core size and of the jet's transverse width as a function 
of core separation, are very important parameters for theoretical jet models. The class of models,
which are based on the magnetic sling shot mechanism (Blandford \& Payne 1982) predict a minimum
transverse width of the jet after initial acceleration to be of order $\geq \gamma \cdot 2R_L$, 
(with bulk Lorentz factor $\gamma$ and light cylinder radius $R_L$,  e.g. Tomimatsu \& Takahashi 2003). 
Notably such jets are expected to be hollow (edge-brightened), just as observed here.
After collimation, the light cylinder has a radius of at least $>(10-50)$ $R_{S}$ 
(Camenzind 1990, Fendt \& Memola, 2001, Casse \& Keppens 2004), which sets a lower bound to the observable
jet width. With jet Lorentz-factors in the range $3 \leq \gamma \leq 6$ (Biretta et al. 1999),
we expect a minimum jet width of $\geq (60 - 600) R_{S}$. 
The observed diameter of the jet (at $0.5 \leq r \leq 2$\,mas) is $2R_L= (0.5-0.7)$\,mas, 
or $(140 - 200) R_{S}$ (see Figure 1). This is at the lower end of the expected range, 
favoring a moderately low bulk Lorentz-factor ($\gamma < 2-3$) in this region.

At its origin near the black hole, the diameter of the jet depends on the size of the region,
where it is initially accelerated. In the Blandford \& Payne type models, the twisted magnetic field lines
are anchored in the inner part of the rotating accretion disk. The minimum jet width is then determined
by the diameter of the least stable orbit, which is at $6 R_{S}$ for a Schwarzschild BH, and  $2 R_{S}$
for a rotating Kerr BH (e.g. Meier et al. 2001).
The observed small size of the VLBI core ($<15 R_{S}$) is already within a factor of $2-3$ near these values,
suggesting that the width of the `jet-nozzle' is indeed of order of the light cylinder diameter.
Since the radiating jet boundary cannot be indefinitely thin, one could expect that the observed jet width is 
a bit larger than the light cylinder. In this case, the observed small size of the VLBI core would point 
more towards jet models, in which the jet gains energy directly from BH-rotation (Blandford \& Znajek 1977,
see also McKinney 2006). Better VLBI images and more specific 
theoretical modeling of jet opening angle and speed will help to discriminate between the
different types of models (e.g. Blandford \& Payne versus Blandford \& Znajek),
and may tell to what degree black hole rotation plays a role in jet formation.


\begin{figure}[h]
\begin{minipage}{13pc}
\includegraphics[width=13pc,angle=-90]{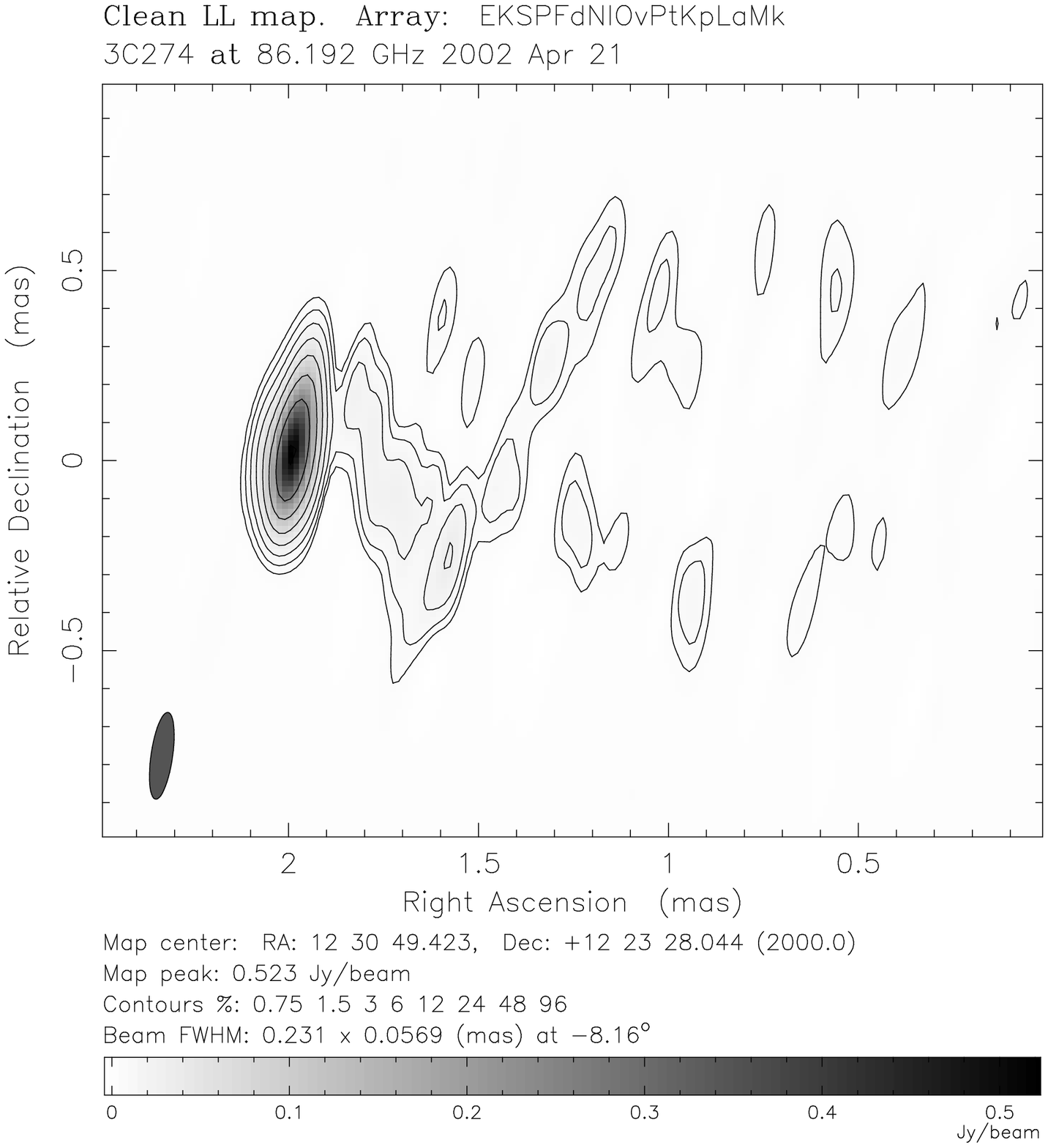}
\end{minipage}
\begin{minipage}{13pc}
\includegraphics[width=13pc,angle=-90]{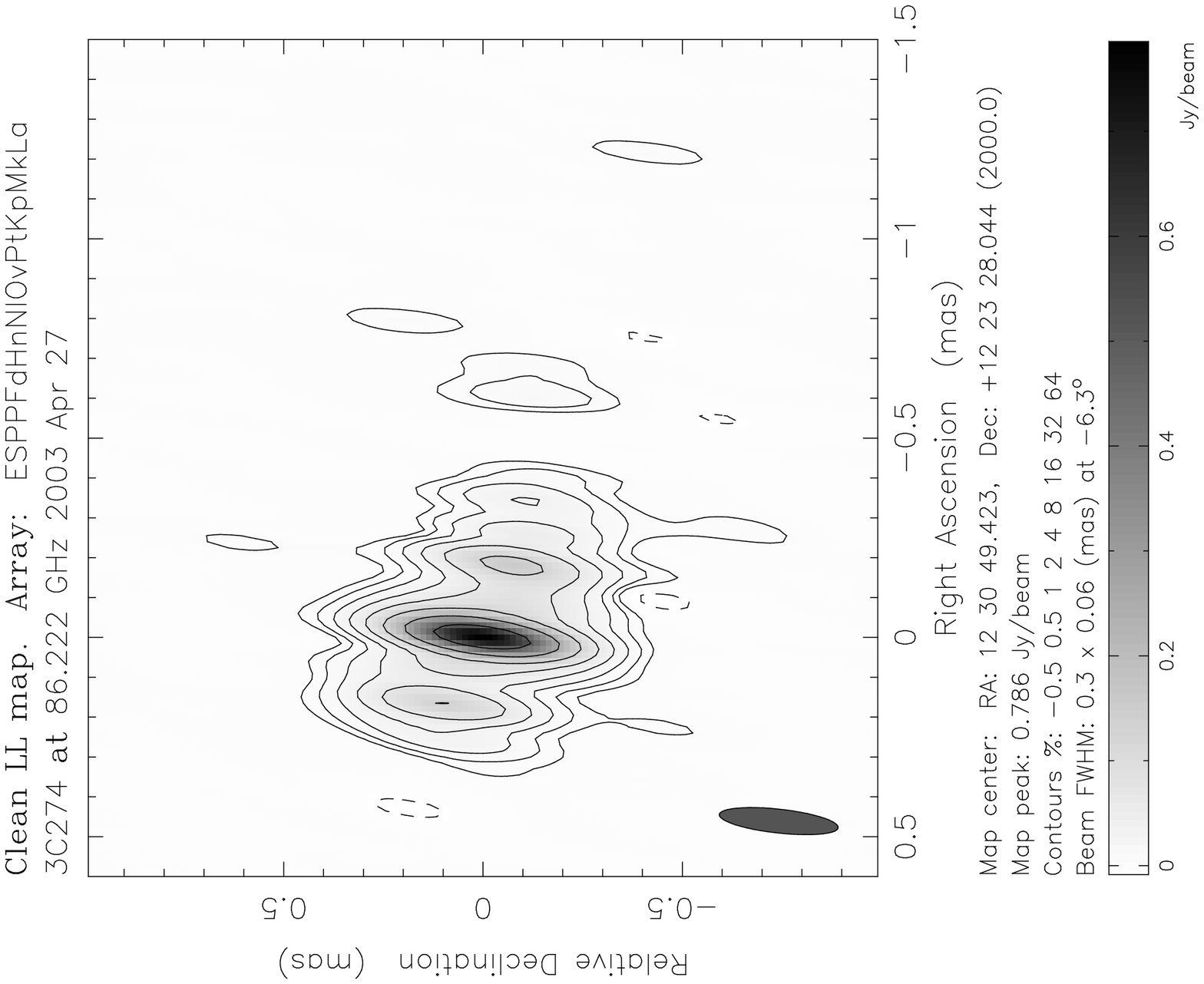}
\end{minipage}
\hspace{-2pc}
\begin{minipage}{13pc}
\includegraphics[width=13pc,angle=-90]{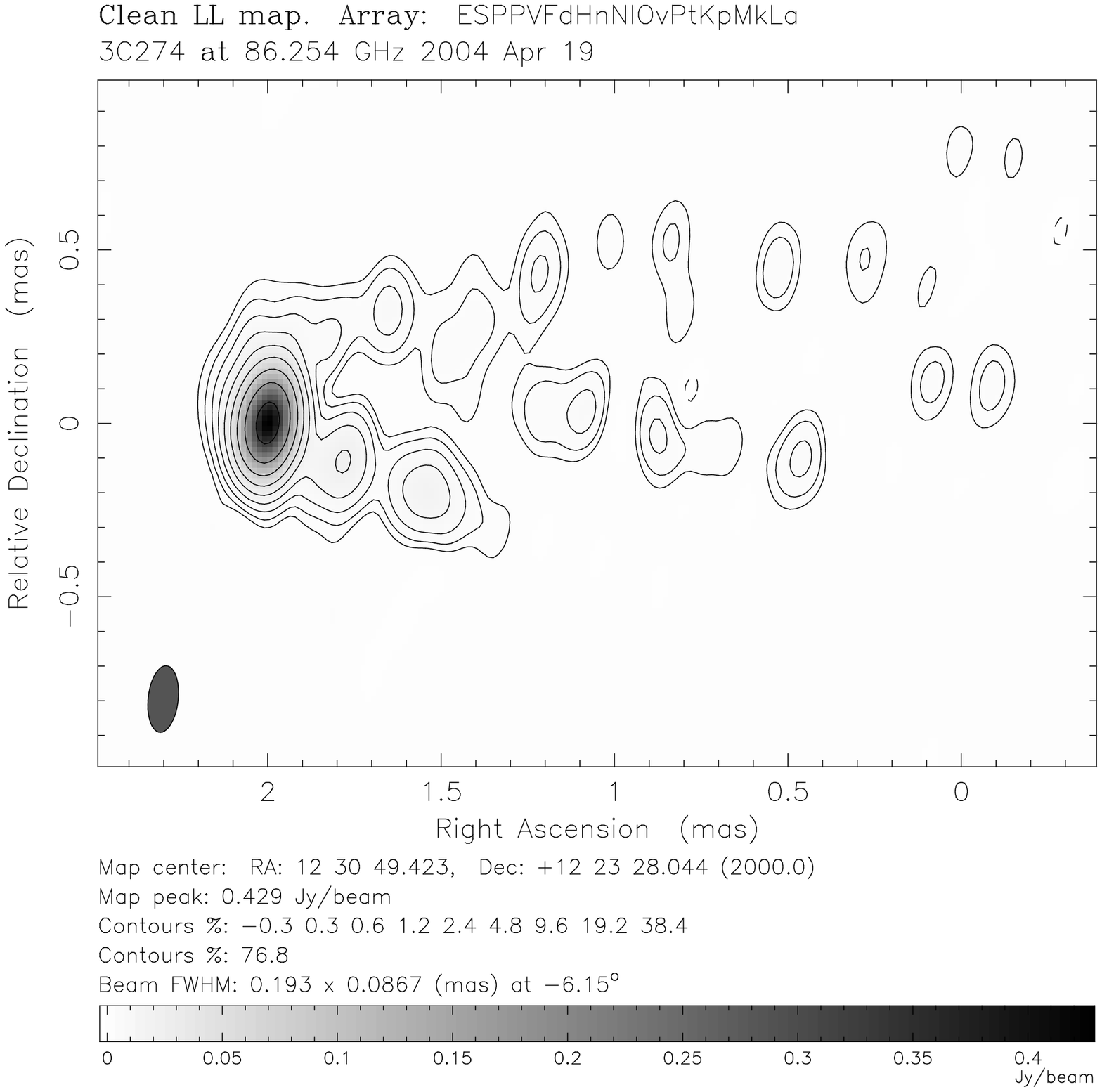}
\end{minipage}
\caption{\label{m87maps}86\,GHz VLBI maps of the jet in M\,87. The observations were performed
(from left to right) in April 2002, 2003, and 2004. The maps result from global VLBI observations using the GMVA, 
which combines European radio telescopes with the VLBA.}
\end{figure}

\begin{figure}[t]
\begin{minipage}{18pc}
\includegraphics[width=18pc,angle=-90]{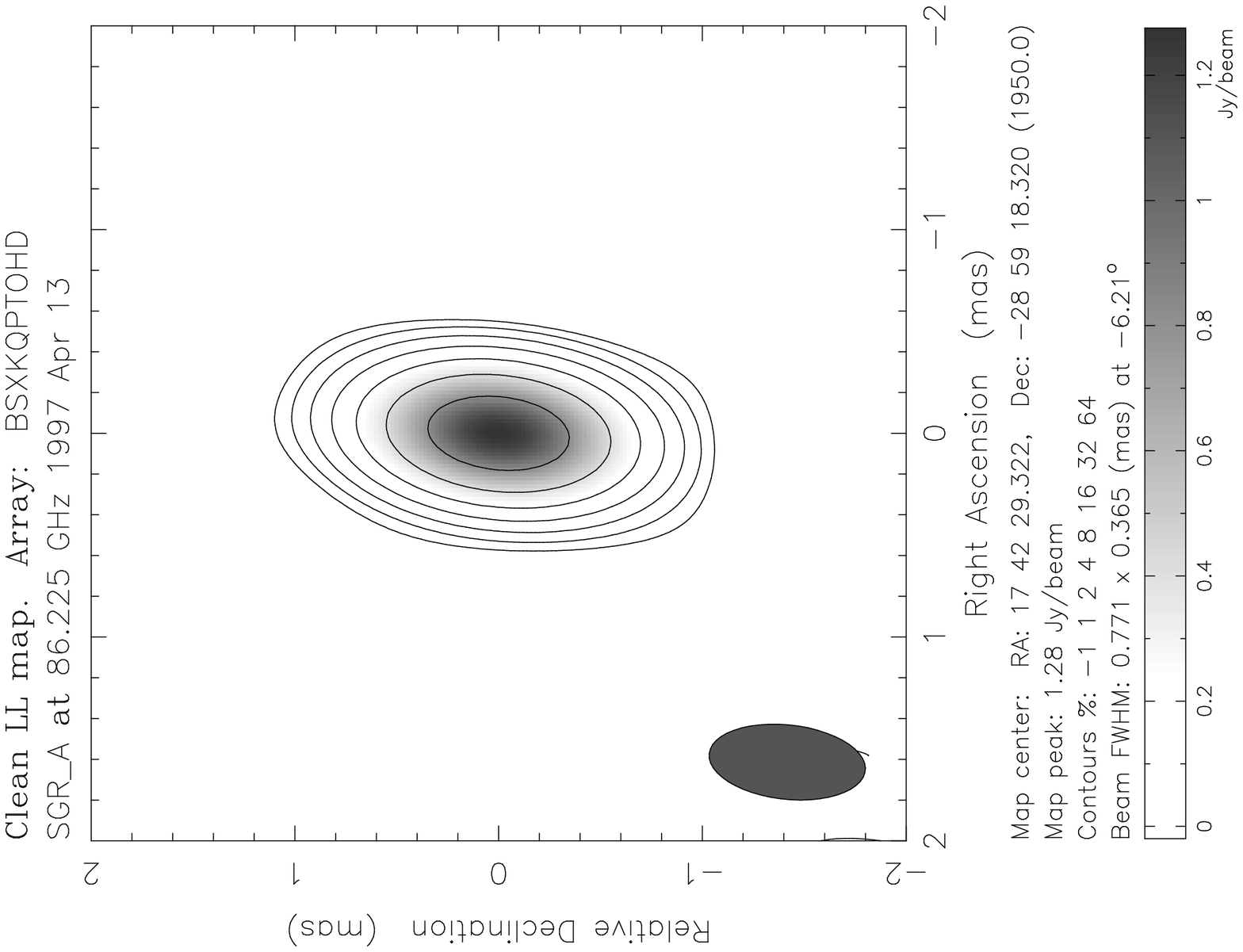}
\caption{\label{sgra97}86\,GHz VLBI image of Sgr\,A* observed in April 1997. 
The participating antennas are: Effelsberg (100\,m), Pico Veleta (30\,m), Haystack (37\,m), Quabbin (14\,m), Pie Town (25\,m), 
Kitt Peak (12\,m), Owens Valley (5 x 10\,m). A circular Gaussian parameterizes Sgr\,A* with
flux density $S=1.73 \pm 0.25$\,Jy and size $\theta=0.28\pm0.08$\,mas.}
\end{minipage}
\hspace{1pc}
\begin{minipage}{18pc}
\includegraphics[width=13pc,angle=-90]{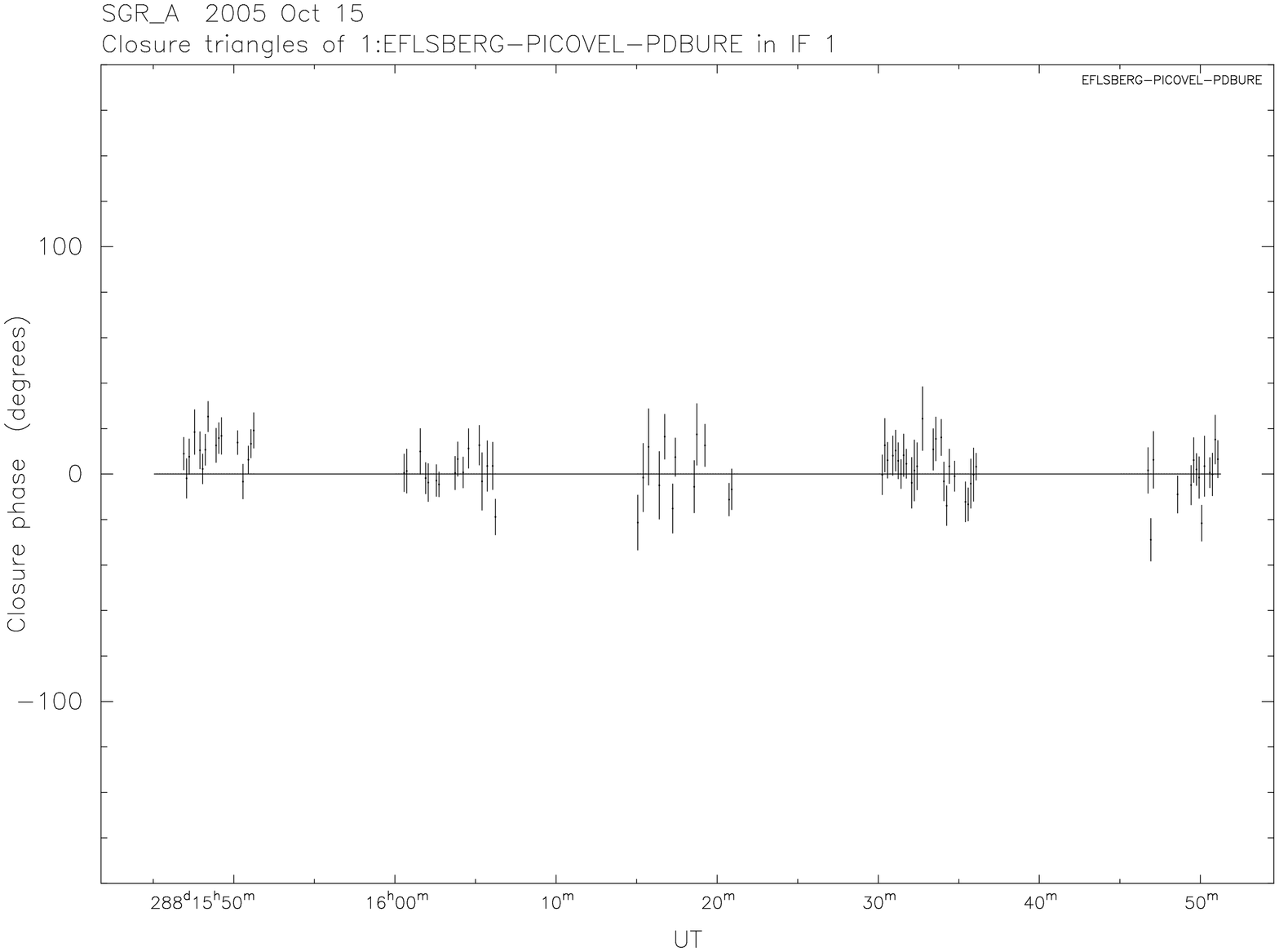}
\caption{\label{3mmclo}Closure phase of Sgr\,A* plotted versus time. Data are from a 3 station VLBI experiment performed 
in October 2005 at 86\,GHz using the telescopes at Effelsberg, Pico Veleta and Plateau de Bure (phased). The closure phase
is zero within $\pm 10^\circ$, suggesting a point-like or almost perfectly symmetrical brightness distribution.}
\end{minipage}
\end{figure}

\begin{figure}
\begin{minipage}{24pc}
\includegraphics[width=18pc,angle=-90,bb=64 39 567 705,clip=]{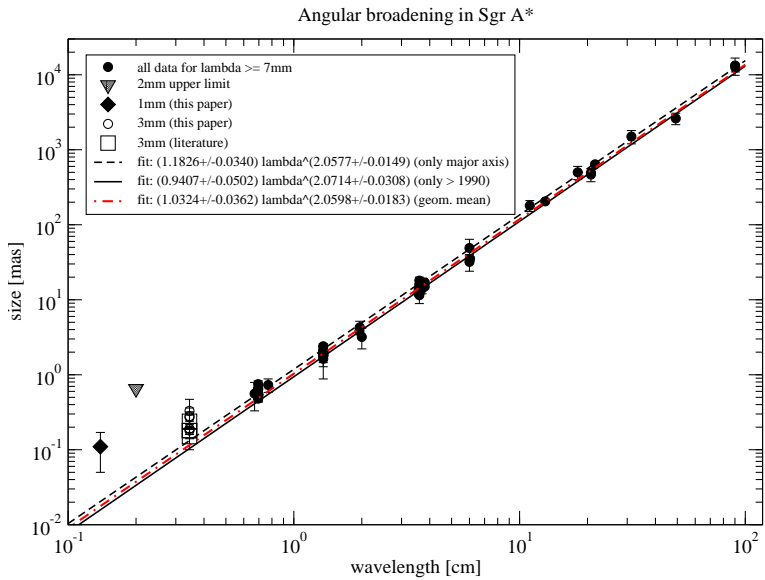}
\end{minipage}
\hspace{1pc}
\begin{minipage}[h]{12pc}
\caption{\label{size}Angular size of Sgr\,A* plotted versus wavelength. Lines denote fits ($a \lambda^\beta$), 
excluding the short wavelengths $\lambda \leq 7$\,mm, which are affected by interstellar scattering. The fit parameters $a$, $\beta$
are given inside the figure. Published 3\,mm measurements are plotted as open squares, new 3\,mm data (this paper) by
open circles. New measurements at 2\,mm (shaded triangle, an upper limit) and 1.4\,mm (diamond) are also shown.}
\end{minipage}
\end{figure}

\section{VLBI observations of Sgr\,A* at 3, 2, and 1\,mm wavelength}

{\bf 3\,mm:}
From the early 1990's onwards, Sgr\,A* was repeatedly observed with VLBI at 3.5\,mm (86\,GHz) in various array configurations,
using European, American and global VLBI arrays (Krichbaum et al. 1994, Rogers et al. 1994, Doeleman et al. 2001, etc.). 
In Figure 2 we show one of the early maps from global 3\,mm VLBI, which reveals a partially resolved point sources,
similar to the recently published VLBA image (Shen et al. 2005). At mm-wavelengths, the accurate determination of flux density and size 
is often affected by calibration uncertainties (low source elevation on the northern hemisphere, 
atmospheric opacity variations). This requires special observing and calibration methods.
In parallel to the ongoing research with the VLBA, we have continued to observed Sgr\,A*, using the sensitive
and large antennas at MPIfR (Effelsberg 100\,m telescope) and at IRAM (Pico Veleta 30\,m, Plateau de Bure 6 x 15\,m).
With baseline lengths of $\sim 750 - 1700$\,km, and a detection sensitivity of $\gsim 50$\,mJy ($7\sigma$, 512 Mbit/sec)
per baseline, this array measures the visibility amplitudes and (closure-) phases with particular high $\rm{SNR}$.
In April and October 2004,  Sgr\,A* was moderately bright with flux densities of $2.40 \pm 0.45$\,Jy and $2.10\pm 0.23$\,Jy,
respectively. In October 2005, however, the source was detected with a much higher signal-to-noise ratio than before
($\rm{SNR} = 96$ on PV-PdB).  A preliminary analysis
of the new data yields (via circular Gaussian model fitting) a flux density of $S=3.6\pm0.5$\,Jy and a size of $\theta=0.27\pm0.05$\,mas.
The relatively high source flux is confirmed independently by ATCA measurements, performed 3 days after our VLBI experiment
($S=2.0-3.5$\,Jy, Z.-Q. Shen, priv. comm.). The source size appears marginally larger (by $2-3\sigma$), when compared to our previous size 
measurements in 2004 ($\theta=0.19\pm 0.05$\,mas and $0.18\pm0.01$\,mas), and in comparison  to
the VLBA measurements of Shen et al. 2005, who obtained $\theta=0.18\pm 0.02$\,mas with the VLBA. 
As in previous observations, the closure phase (see Figure 3) is close to zero (within $\pm 10^\circ$),
showing no evidence for a deviation from the known point-like (or symmetric) structure. 
In Figure 5, we plot all available 86\,GHz VLBI size measurements, using the data from the literature and including 
own unpublished old and new data. If one excludes the new
data point of October 2005, the weighted mean average size is $0.18\pm0.01$\,mas. This appears significantly smaller
than the October 2005 measurement. For comparison we also show in Fig. 5 the sizes at 43\,GHz (weighted mean: $0.55\pm0.03$\,mas) and the 
flux density of Sgr\,A* at 86\,GHz. It will need a denser 3\,mm VLBI monitoring to check, if 
the brightness distribution of Sgr\,A* indeed is variable. In view of the observed flux density variations at millimeter wavelengths
(see Miyoshi et al., Miyazaki et al., Marrone et al., this conference), structural variability of Sgr\,A* would be not 
unexpected (see Bower et al. 2004 for possible size variations  at 7\,mm).

{\bf 2\,mm:}
VLBI observations at wavelengths shorter than 3\,mm are still difficult and are limited 
by the number of available radio telescopes and their sensitivity. In a 
pilot VLBI experiment at 147\,GHz (2\,mm), which aimed at first fringe detection across the Atlantic at this short wavelength,
Sgr\,A* was observed among a dozen other AGN (Krichbaum et al. 2002). Several AGN
were detected on the long ($\sim 8500$\,km) baseline between Spain (Pico Veleta) and the Arizona telescopes 
(Kitt Peak (KP) 12\,m, Heinrich-Hertz telescope (HHT) 10\,m), which limits the size of their compact emission regions 
to $\leq 25-40$\,$\mu$as. Sgr\,A* was detected with a signal-to-noise ratio of $\sim 7$ on the short (85\,M$\lambda$) 
baseline between KP and HHT. With a total flux density of $\sim 1.9$\,Jy and a correlated flux density 
of $S_c= 1.6 \pm 0.25$\,Jy an upper limit to its size of $\leq 0.7$\,mas
can be derived, fully consistent with the measured sizes at 3\,mm and 1.4\,mm. 

\begin{figure}[h]
\begin{minipage}{18pc}
\includegraphics[width=13pc,angle=-90]{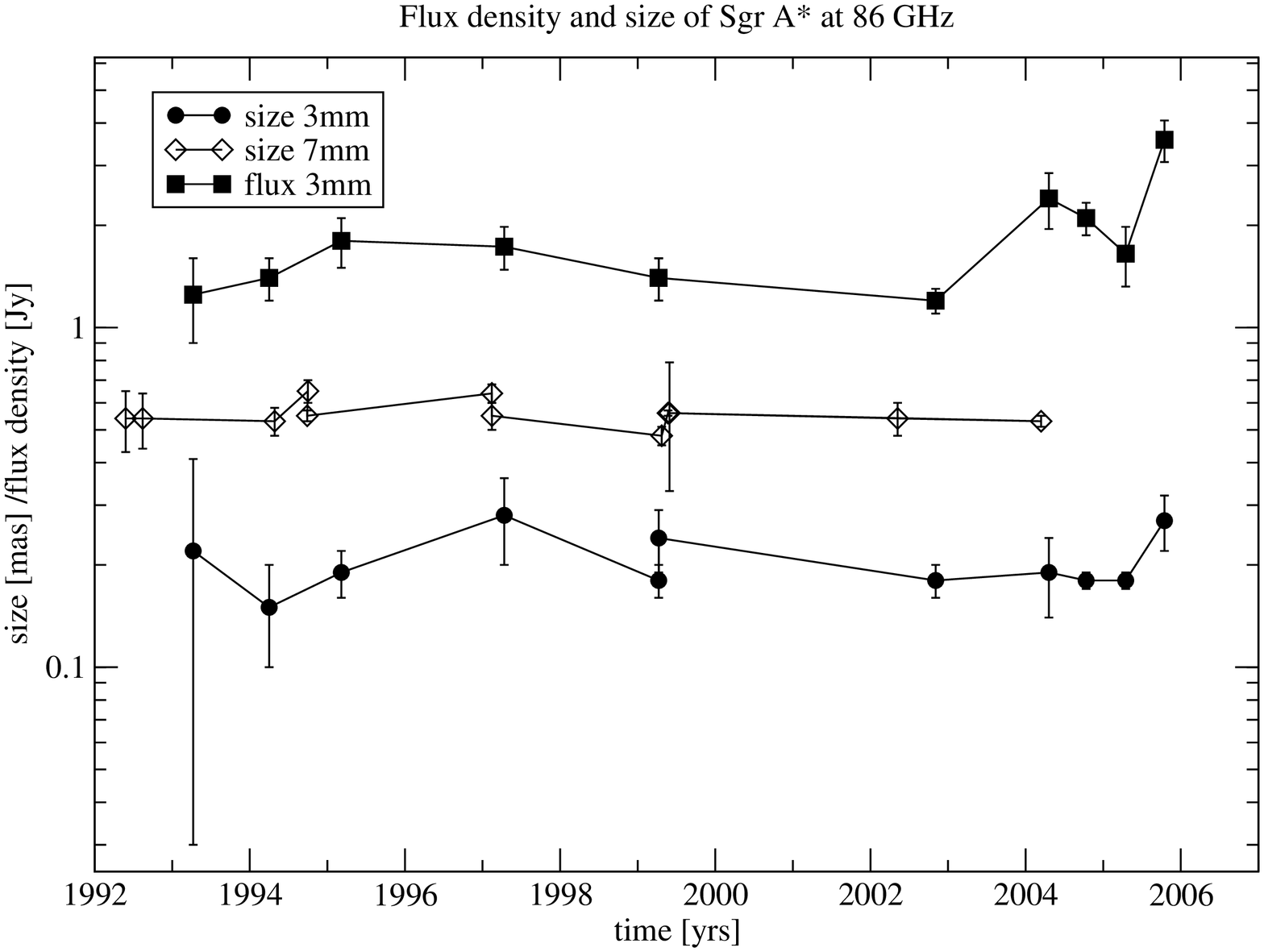}
\caption{\label{sizetime} Measured VLBI sizes of Sgr\,A* at 86\,GHz (filled circles) and at 43\,GHz (open diamonds) 
plotted versus time. Data are taken from the literature, including our new and unpublished 86\,GHz data.
All sizes were calculated from the geometric mean of major and minor axis. We also show the VLBI flux density
(filled squares) at 86\,GHz.
}
\end{minipage}
\hspace{1pc}
\begin{minipage}{18pc}
\includegraphics[width=12pc,angle=-90]{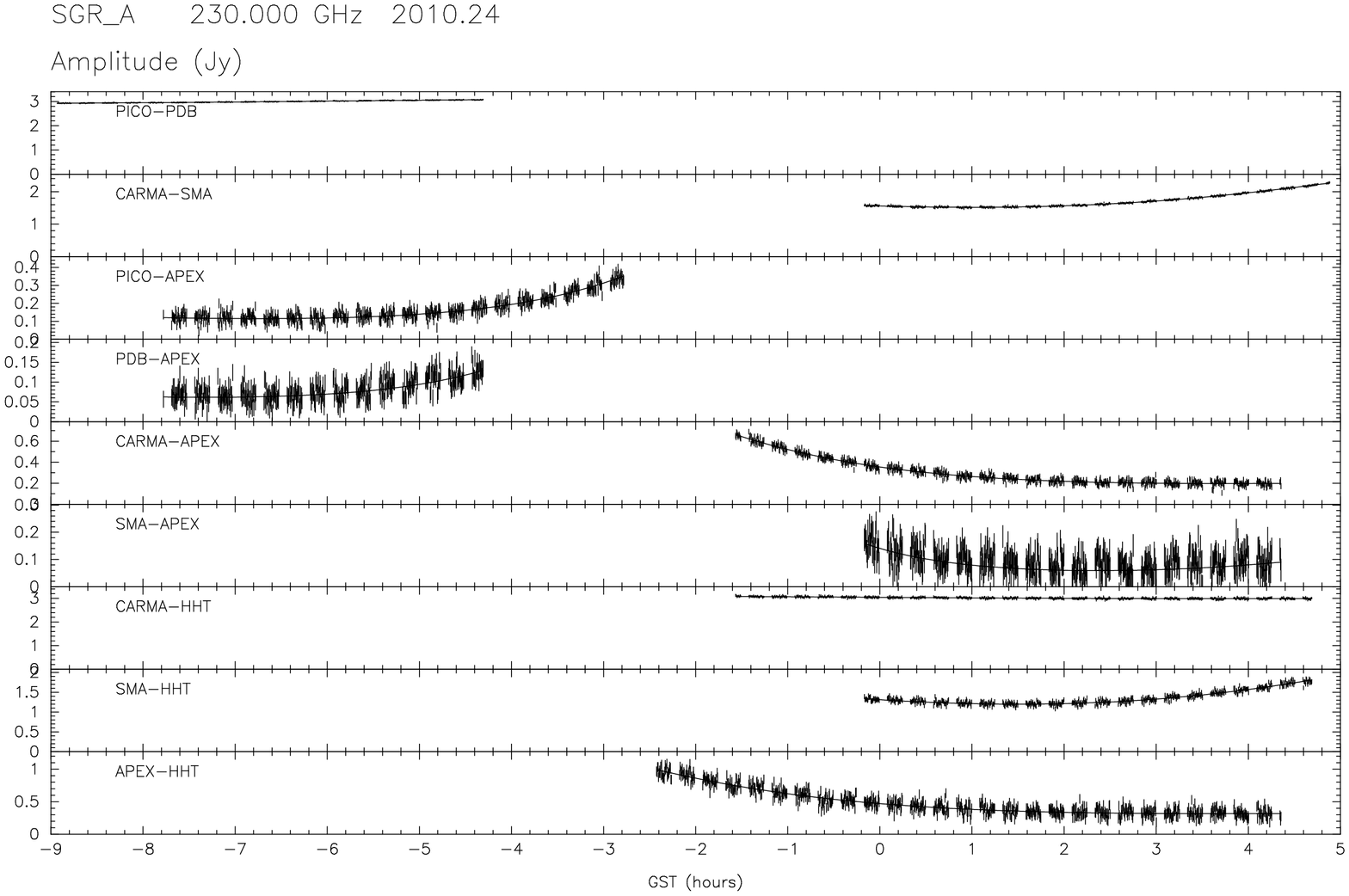}
\includegraphics[width=12pc,angle=-90]{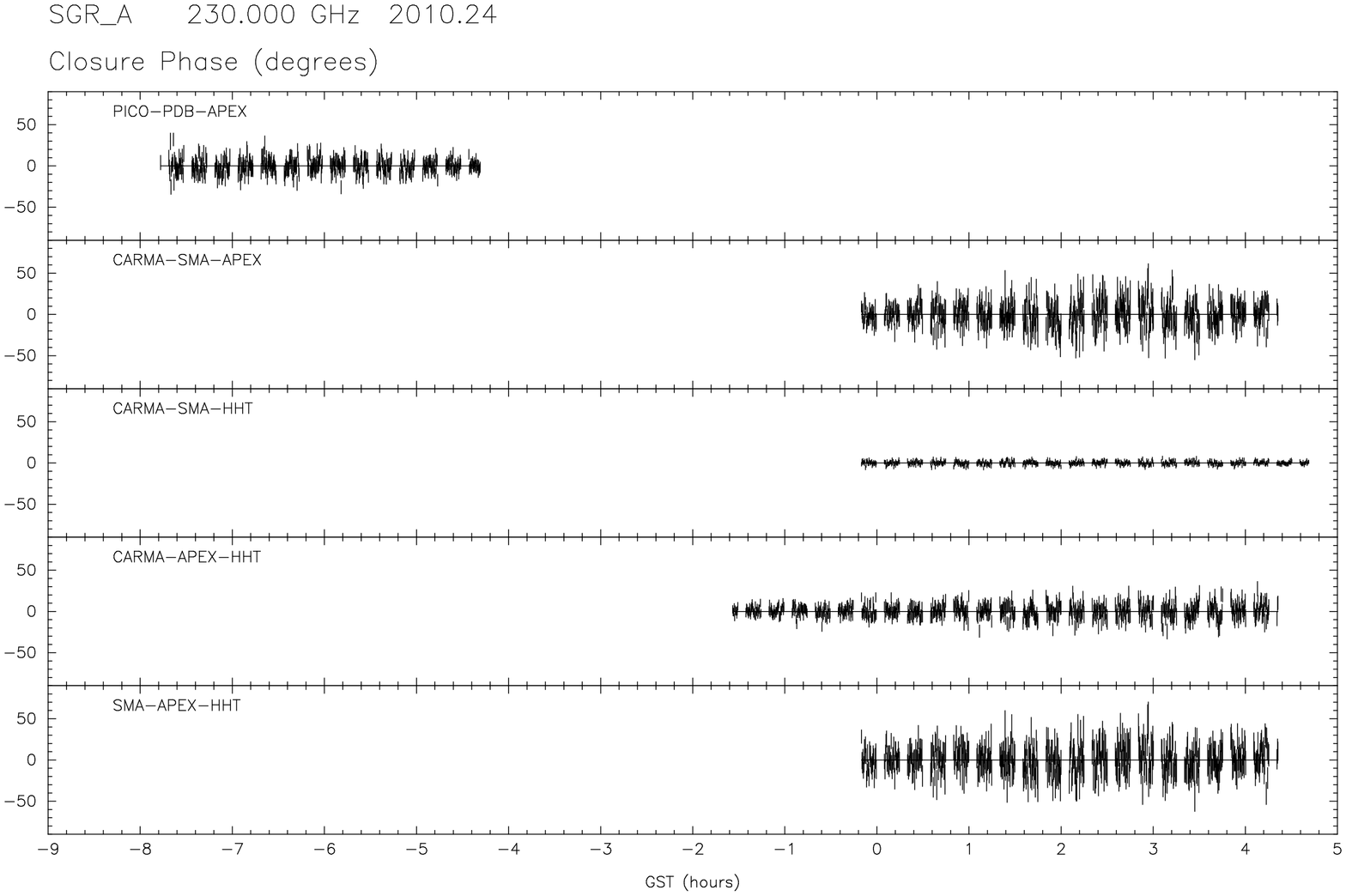}
\end{minipage}
\caption{\label{1mmvis} Right: Simulated visibility amplitudes (top) and closure phase (bottom) for a future 1.3\,mm VLBI 
experiment on Sgr\,A* with the following telescopes: Pico Veleta (Spain), Plateau de Bure (France), CARMA (California), SMA
(Hawaii), HHT (Arizona) and APEX (Chile). For the simulation, a circular Gaussian of 2.5\,Jy and $\sim 25$\,$\mu$as FWHM size was assumed.
}
\end{figure}

{\bf 1\,mm:}
In April 2003, our attempt to observe Sgr\,A* with global VLBI at 230\,GHz failed due to weather and technical problems.
In this experiment, which was motivated in a similar manner as the 2\,mm VLBI experiment described before, 
several bright AGN were observed. Finally two blazars (3C\,454.3, 0716+714) 
were detected with $\rm{SNR} \gsim 7$ on the 6.4\,G$\lambda$ baseline between Pico Veleta and HHT (Krichbaum et al. 2004). 
Although limited by receiver performance and stability of local oscillators at HHT, this experiment
demonstrates the technical feasibility of global VLBI at 1.3\,mm, and sets a new record in angular
resolution. For the detected compact radio sources we obtained sizes of $\lsim 30$\,$\mu$as. With such a 
high angular resolution, which would correspond to a spatial resolution of $\sim 3$ Schwarzschild 
radii in Sgr\,A*, the direct (VLBI-) imaging of the immediate vicinity of a SMBH should become possible  
within the next few years. 

We now like to make a remark with regard to the previous 215\,GHz VLBI observations of Sgr\,A* published
earlier (Krichbaum et al. 1998).
Recently, Eckart et al. 2006 (also Eckart et al., this conference) detect 
an emission component D1 of $\sim 170$\,mas size, located $80 \pm 50$ mas 
west of Sgr A*. With single dish radio telescopes operating at short mm-/sub-mm wavelengths, in-beam confusion between Sgr\,A* and
D1 leads to an over-estimate of the total flux density of Sgr\,A* in this spectral band 
(by $\Delta S_{215\,\rm{GHz}} \simeq 0.75$\,Jy for $T \sim 10^3$\,K), at least partially removing 
the so called sub-mm excess (e.g.. Serabyn et al. 1997,  Falcke et al. 1998). In fact, a less pronounced or 
non-existing sub-mm excess, leads to a flat synchrotron-self-absorbed radio spectrum as observed in many other AGN.
This then would allow an interpretation of the high energy data via the 
synchrotron-self-Compton process (see Eckart et al. 2004 \& 2006).

The existence of such an extended emission region, which is fully resolved and therefore invisible to VLBI,
was considered by Krichbaum et al. 1998 as a possibility to explain the apparent discrepancy between observed
total flux (from single antenna measurements) and the correlated VLBI flux, which originates from the compact
component. Relating the sub-mm excess of the total spectrum mainly to the flux contribution of component D1, allows
to redetermine the total flux of Sgr\,A* at 215\,GHz and reject `case 1' in the discussion of Krichbaum et al. 1998. 
This leads to refined estimate of the observed size for Sgr\,A* at 215\,GHz: 
$\theta = (110 \pm 60) \mu$as, which corresponds to a linear size of $R_{S} = 11 \pm 6 $ Schwarzschild radii
and a robust lower limit of $R_{S} \geq 5$.

\section{Intrinsic Source Size}
The observed frequency dependence of the source size of Sgr\,A* (Fig. 4) is commonly interpreted as being due to scatter broadening
by the intervening interstellar medium. The intrinsic size
is determined by the deconvolution of the observed size $\theta_{\rm obs}$  with the extrapolated scattering size
$\theta_{\rm scat}$: $\theta_{\rm int} = \sqrt{\theta_{\rm obs}^2 - \theta_{\rm scat}^2}$. The exact form of the assumed
scattering law ($\theta_{\rm scat} = a \lambda^\beta$, with fit parameters $a,\beta$) determines $\theta_{\rm int}$.
In Figure 4 we show a few illustrative least-square fit lines, which were obtained from all at present available size measurements,
selecting those above $\lambda \geq 1.3$\,cm\footnote{these wavelengths should be least affected by a finite intrinsic source size}
and performing a fit to (i) only the major axis, (ii) the geometric mean of major and minor axis, and (iii) the most recent 
($> 1990$) data. The resulting mean slopes in the range of $\beta=2.06 \pm 0.01$ deviate slightly from the slope of $\beta=2$
other authors recently have used (Bower et al. 2004, Shen et al. 2005 and references therein). 
Depending on the number of data points included in the fit, their weighting scheme and the frequency range over 
which the fit is performed, one obtains slightly different estimates of the intrinsic size of Sgr\,A*. Since the size
determination at the shortest wavelengths ($\lambda \leq 3$\,mm) is still restricted to a circular Gaussian component,
we determine the de-convolved intrinsic size of Sgr\,A* using the geometric mean size from major and minor axis at \underline{all}
wavelengths. Formally we obtain the following 
sizes\footnote{For $d= 8.0$\,kpc and $M_{\rm BH}= 4 \cdot 10^6M\solar$ an angle of 0.1\,mas corresponds to $1 R_{S}$}: 
$25 \pm 7 R_{S}$ at 7\,mm, $15 \pm 3 R_{S}$ at 3.4\,mm, $< 65 R_{S}$ at 2\,mm, and $11 \pm 6 R_{S}$ at 1.4\,mm.

\section{Future Outlook}
Global mm-VLBI performed at the shortest wavelengths ($\leq 3$\,mm) will allow to study Sgr\,A* (and
other AGN like, e.g. M\,87) with a resolution of only a few Schwarzschild radii, facilitating the direct imaging of
the immediate vicinity of super-massive black holes. Global mm-VLBI observations of Sgr\,A* are clearly a great challenge and demand
coordinated efforts.
In Figure 6 we show simulations of visibility amplitudes and closure phases for a realistic near future global mm-VLBI observation 
of Sgr\,A* at 230\,GHz. New telescopes like APEX (to be later replaced by ALMA) located in the southern hemisphere,
and phased interferometers operating as single VLBI antennas (PdB, CARMA, SMA) are crucial to provide
the necessary sensitivity on the longest baselines. When these telescopes become available for mm-VLBI,
we may hope to image the environment of the SMBH in Sgr\,A* and
detect possible deviations from circular symmetry, which are theoretically predicted for a
rotating accretion disk black hole system.


\end{document}